\documentclass[aps,prl,twocolumn,showpacs,preprintnumbers,superscriptaddress]{revtex4}
\usepackage{mathrsfs}
\usepackage{amssymb}
\usepackage{amsmath}
\usepackage{graphicx}
\usepackage{dcolumn}
\usepackage{bm}
\usepackage{graphicx}
\usepackage{float}
\usepackage[normalem]{ulem}
\usepackage{color}
\usepackage{tikz-feynman}
\usepackage{slashed}
\usepackage{caption}

\begin{document}
\title{Constraining New Muonic  Interactions Meditated by Axion-Like-Particles}
\author{H. Yan}
\email[Corresponding author: ]{hyan@caep.cn}\affiliation{Key Laboratory of Neutron Physics, Institute of Nuclear Physics and Chemistry, CAEP, Mianyang, Sichuan, 621900,China }

\author{G.A. Sun}
\affiliation{Key Laboratory of Neutron Physics, Institute of Nuclear Physics and Chemistry, CAEP, Mianyang, Sichuan, 621900,China }

\author{S.M. Peng}
\affiliation{Institute of Nuclear Physics and Chemistry,CAEP, Mianyang, Sichuan, 621900,China }

\author{H. Guo}
\affiliation{Department of Physics, Southeast University, Nanjing, 211189, China }

\author{B.Q. Liu}
\affiliation{Key Laboratory of Neutron Physics, Institute of Nuclear Physics and Chemistry, CAEP, Mianyang, Sichuan, 621900,China }

\author{M. Peng}
\affiliation{Key Laboratory of Neutron Physics, Institute of Nuclear Physics and Chemistry, CAEP, Mianyang, Sichuan, 621900,China }

\author{H. Zheng}
\affiliation{Key Laboratory of Neutron Physics, Institute of Nuclear Physics and Chemistry, CAEP, Mianyang, Sichuan, 621900,China }

\date{\today}
\begin{abstract}
ALPs (Axion Like Particle) beyond the standard model are solutions to several important problems of modern physics. One way to detect these particles is to detect the new interactions they meditate. Many experiments have been
 performed to search for  these new interactions in ranges from $\sim\mu$m to astrophysical range. At present, nearly all known experiments searching for the ALP-meditated long range new interactions use sources or probes containing protons, neutrons and electrons. Constraints for other fermions such as muons   are scarce, though muons might be the most suspicious particles which could take part in new interactions, considering their involvement of several well known puzzles of modern physics. In this work, we discuss the possibility of explaining the anomalous magnetic moment of muons by the long range muonic new interactions mediated by ALPs. We also give a constraint for the scalar-pseudo-scalar(SP) type interaction meditated by muonic ALPs. We propose to further search the muonic SP type interaction by muon spin rotation experiments.

\end{abstract}
\maketitle
\section*{Introduction}
Suggested solutions for several important problems of modern physics have led to new interactions meditated by new particles.  Among the proposed new particles, Axions, or ALPs(Axion Like Particles) are particularly interesting.  The new particles are supposed to be light and only couple very weakly to the ordinary matter.  On one hand, axions are possible candidates for the dark matter which remains to be one of the most important unsolved problems in both particle physics and astrophysics \cite{PDG16,FU17}.  On the other hand, axions have attracted a lot of attention in high energy physics since they probably provide the most promising solution to preserve the CP-symmetry in strong interactions \cite{KIM10}. In fact, the axion was originally introduced to solve the strong CP problem in QCD in which new bosons emerge as a consequence of the spontaneous breaking of Pecci-Quinn symmetry \cite{PEC77,MOO84,PEC06,MAN14}.
It is very difficult to search for Axions or ALPs in laboratory and they have eluded detection so far. Since macroscopic interactions can be meditated by the new light bosons, a possible method to search for them is to probe the boson field generated by a macroscopic body. Recently, various experiments have been performed or proposed to search for new interactions involved with the coupling to spins of neutrons/electrons \cite{VAS09,HOE11,RAF12, BUL13,LED13,TUL13, LES14,  YAN14, CHU15, KOT15, TER15,CHU16,FIC17}. 

The ALPs, if exist, might induce interactions between fermions via the coupling\cite{MOO84}:
\begin{equation}
 \mathcal{L}_{I}=\bar{\psi}(g_{s}+ig_{p}\gamma_{5})\psi\phi.
 \end{equation}
 There could be monopole-monopole, dipole-dipole and  monopole-dipole interactions, originated from the SS (Scalar-Scalar),  PP (Pseudo-scalar-Pseudo-scalar) and SP (Scalar-Pseudo-scalar) coupling, respectively. The SP interaction or the monopole-dipole interaction
has attracted much scientific interest recently. The interaction between polarized and unpolarized fermions can be expressed as\cite{MOO84,DOB06}:
\begin{equation}\label{eqnSP}
V_{SP}(r)=\frac{\hbar^{2}g_{S}g_{P}}{8\pi m}(\frac{1}{\lambda r}+\frac{1}{r^{2}})\exp{(-r/\lambda)}\vec{\sigma}\cdot\hat{r}
\end{equation}
where $\lambda=\hbar/m_{\phi}c$ is the interaction range, $m_{\phi}$ the mass of the new scalar boson, $\vec{s}=\hbar\vec{\sigma}/2$  the spin of the
polarized electron, $m$ the fermion mass and $r$ the distance between interacting particles. 
For an unpolarized mass-source as a plane plate, due to the new interaction, the polarized fermions could experience a  pseudo-magnetic field as \cite{YAN14}:
\begin{equation}\label{eqnPB}
\vec{B}_{SP}=\frac{1}{\gamma}\frac{\hbar g_{S}g_{P}}{2m}\rho\lambda e^{-\frac{\Delta y}{\lambda}}[1-e^{-\frac{d}{\lambda}}]\hat{y}
\end{equation}
where $\Delta y>0$ is the distance between the probe and the sample surface, $\hat{y}$ the sample surface normal vector, $d$ the sample thickness, $\rho$ the fermion number density
of the sample, and $\gamma$ the gyromagnetic ratio of the probing particle. According to Eqn.(\ref{eqnPB}),  the exponential factor $e^{-\frac{\Delta y}{\lambda}}$ determines the shortest detection range of the interactions, since practically the source thickness is usually not a problem. 
To detect the new interaction in ranges around $\sim\mu$m, either the source or the probe particle has to be spin polarized, and they have to be close enough.

At present, usually a mass-source containing nuclei, electrons is used for detecting the new interactions. Polarized neutrons or electrons are used as probe to detect the new spin dependent interactions. 
Polarized neutrons or electrons are used as probes to detect the new spin-dependent interactions. However, studies of the long-range new interactions meditated by ALPs for other fermions are scarce. Muons are probably the most suspicious fermions which new interactions might involve with. The charge radius puzzles of the muonic hydrogen \cite{ANT13} and deuteron \cite{POH16} nucleus are well known examples. Parity-violating muonic forces meditated by new massive gauge bosons of MeV$\sim$GeV have been proposed to solve the proton charge radius puzzle \cite{BAT11,BAR12}.

 To our best knowledge, no studies have been conducted for searching ALP-meditated long range muonic interactions yet. In this work, we show that the anomalous magnetic moment of the muon could be explained by introducing the new muonic interactions meditated by light ALPs. Furthermore, by using the muon's EDM (Electric Dipole Moment), we could establish constraints of the parity-violating monopole-dipole interaction mediated by ALPs for muons.

\section{The Anomalous Magnetic Moment and EDM induced by new interactions}
The leading order contribution to the electromagnetic vertex from the new interactions is shown in Fig.1. The new boson line could induce SS, PP and SP type vertexes.
\begin{center}
\begin{tikzpicture}
\begin{feynman}
\vertex (a);
\vertex [right=of a] (b);
\vertex [above right=of b] (f2);
\vertex [above right=of f2] (f3);
\vertex [below right=of b] (f4);
\vertex [below right=of f4] (f5);
\diagram* {
(a) -- [boson, edge label=\(q\)] (b),
(b) -- [fermion,edge label=\(p'+k\)] (f2),
(f2) -- [fermion,edge label=\(p'\)] (f3),
(f5) -- [fermion,edge label=\(p\)] (f4),
(f4) -- [fermion,edge label=\(p+k\)] (b),
(f4) -- [scalar, out=45, in=-45, edge label'=\(k\)] (f2),

};
\end{feynman}
\end{tikzpicture}\\
\end{center}
{FIG.1 The electromagnetic vertex correction contributed by new interactions.}\\

The new interactions, if exist, would not only change the anomalous magnetic moment of the fundamental fermion, but also induce the EDM which can be measured for many particles \cite{PDG16}. According to Refs.\cite{BER91,ITZ80}, the general formula  corresponding to the Feynman diagram shown in Fig.1 can be expressed as:
\begin{eqnarray*}
\bar{u}(p')\Lambda^\mu u(p)=\bar{u}(p')[\gamma^\mu F_1(q^2)+i\frac{\sigma^{\mu\nu}q_\nu}{2m}F_2(q^2)+\\
\gamma^5\frac{\sigma^{\mu\nu}q_\nu}{2m}F_3(q^2)+\gamma^5(q^2\gamma^\mu-2m\gamma^5q^\mu)F_4(q^2)]u(p)
\end{eqnarray*}
where $q=p'-p$. Here $eF_1(0)$ gives the renormalized charge, and the $F_4$ term is the possible axial current induced by the new interactions which may not necessarily preserve the parity. These two terms are irrelevant in this work. e$F_2(0)$/2m is the anomalous magnetic moment, and -e$F_3(0)$/2m gives the EDM.

Using the well known Gordon decomposition formula and techniques presented in Ref.\cite{ZEE10}, one can derive the anomalous magnetic moment induced by the SS or PP interactions.
If the new boson only meditates the SS interaction , we find that \cite{SUPL}
\begin{eqnarray}
F_2(0)=-\frac{g_Sg_S}{8\pi^2}S(x),
\end{eqnarray}
where $S(x)$ is defined as
\begin{eqnarray*}
S(x)=\frac{3}{2}-x^2+x^2(x^2-3)\ln{x}+x(x^2-1)\sqrt{x^2-4}\times\\
\times[\tanh^{-1}{(\frac{x}{\sqrt{x^2-4}})}-\tanh^{-1}{(\frac{x^2-2}{x\sqrt{x^2-4}})}]
\end{eqnarray*}
with $x={m_{\phi}}/{m}$. 
Similarly, if the new particle only causes the PP interaction,  we have \cite{SUPL}
\begin{eqnarray}
F_2(0)=-\frac{g_Pg_P}{8\pi^2}P(x)
\end{eqnarray}
where  $P(x)$ is defined as
\begin{eqnarray*}
P(x)=\frac{1}{2}+x^2-x^2(x^2-1)\ln{x}+\frac{x^3(3-x^2)}{\sqrt{x^2-4}}\times\\
\times[\tanh^{-1}{(\frac{x}{\sqrt{x^2-4}})}-\tanh^{-1}{(\frac{x^2-2}{x\sqrt{x^2-4}})}].
\end{eqnarray*}

EDM can be induced by the possible new interactions. In this case, to perform the necessary calculations, instead of the Gordon decomposition formula the following identity can be applied
\begin{eqnarray*}
\bar{u}(p')\gamma^5\frac{\sigma^{\mu\nu}q_\nu}{2m_e}F_3(q^2)u(p)=i\bar{u}(p')\gamma^5u(p)\frac{p'^\mu+ p^\mu}{2m_e}F_3(q^2).
\end{eqnarray*}
Implementing similar techniques as before, for the SP type new interaction, we obtain  \cite{SUPL}
\begin{equation}
F_3(0)=\frac{g_Sg_p}{4\pi^2}SP(x)
\end{equation}
where
\begin{eqnarray*}
SP(x)=1-x^2\ln{x}-\frac{x(x^2-2)}{\sqrt{x^2-4}}\times\\
\times[\tanh^{-1}{(\frac{x}{\sqrt{x^2-4}})}-\tanh^{-1}{(\frac{x^2-2}{x\sqrt{x^2-4}})]}.
\end{eqnarray*}
\section{applications to the muon and electron}
There is a $\sim$3.7$\sigma$ \cite{PDG16,BAL16,LIN16} difference between the experimental measurement and the theoretical prediction from the Standard Model for the anomalous magnetic moment $a_\mu$ of the muon. According to Ref.\cite{BLU18}, this difference is explicitly given by
\begin{eqnarray}
1.28\times10^{-9}<|a_\mu^{exp}-a_\mu^{SM}|<4.48\times10^{-9}, 95\% C.L.
\end{eqnarray}
where $a_\mu^{exp}$ is the experimental measured value and $a_\mu^{SM}$ the Standard Model prediction.
If the contribution by new interactions mediated by ALPs is considered,  
 then the difference can be explained.
 The 2-$\sigma$ bands for $g_S^{\mu}g_S^{\mu}$ and $g_P^{\mu}g_P^{\mu}$ can be obtained as shown by the (yellow) shaded regions in Fig.2 and 3. Plugging in the best known muon EDM \cite{BEN09},
\begin{eqnarray}
|d_\mu|<1.8\times10^{-19}e.cm,95\% C.L.
\end{eqnarray}
a constraint for $g_S^{\mu}g_P^{\mu}$ can be established, shown by the (magenta) solid line in Fig.4. It is interesting to notice that $g_S^{\mu}g_S^{\mu}$ and $g_P^{\mu}g_P^{\mu}$ are nonzero while $g_S^{\mu}g_P^{\mu}$  is zero, all at a $\sim10^{-7}$ level.

 \begin{figure}[htbp]\label{Fig.gSgS}
\begin{center}
\includegraphics[scale=0.4, angle=0]{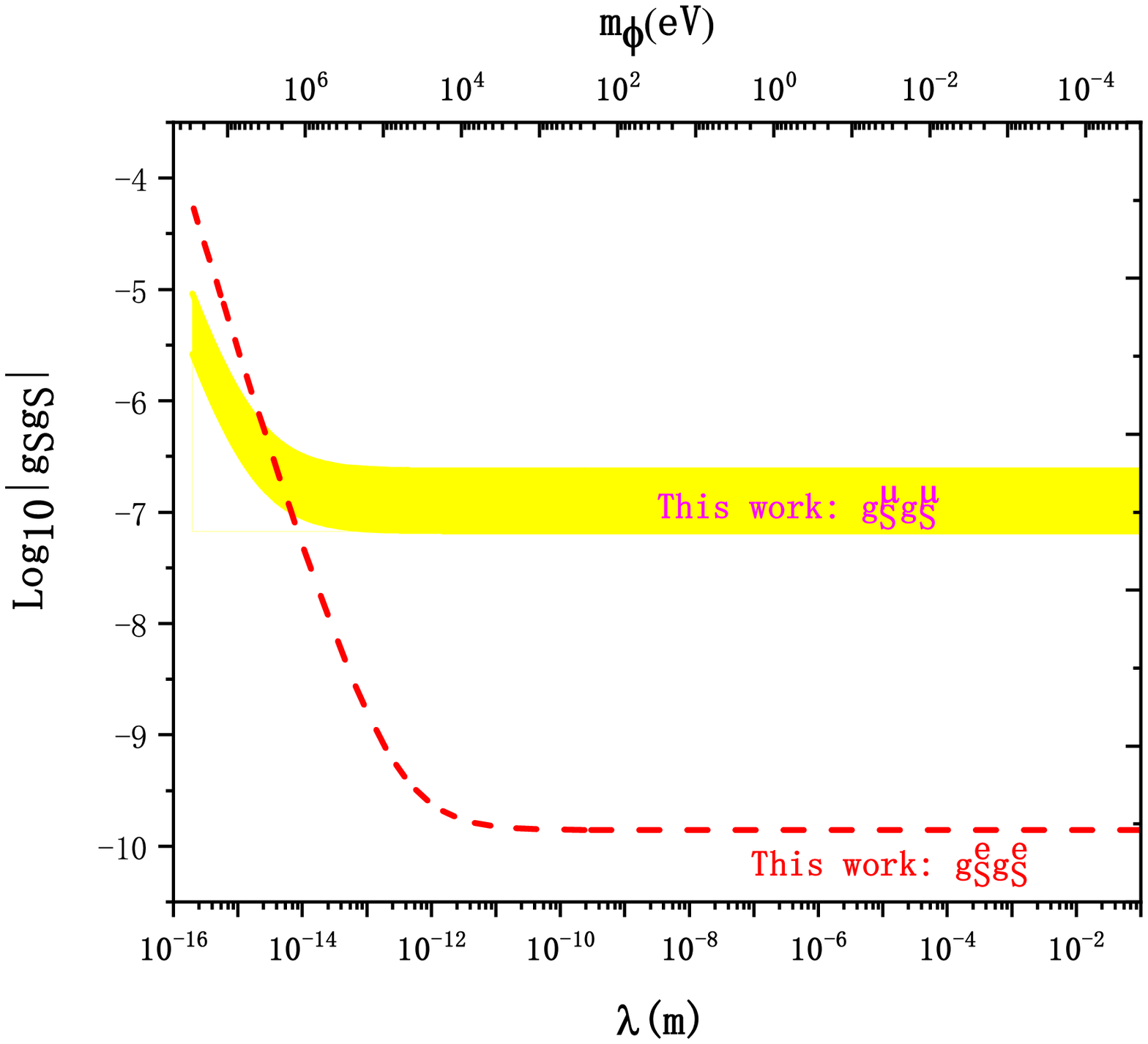}
\end{center}
\centerlast{\small{FIG.2 (Color online) Constraint as as a function of the interaction range $\lambda$ (ALP mass). The (yellow) shaded region is the 2-$\sigma$ band for $|g_S^{\mu}g_S^{\mu}|$, and the (red) dashed line is the constraint to  $|g^e_{S}g^e_{S}|$. }}
\end{figure}

\begin{figure}[htbp]\label{Fig.gPgP}
\centering
\includegraphics[scale=0.4, angle=0]{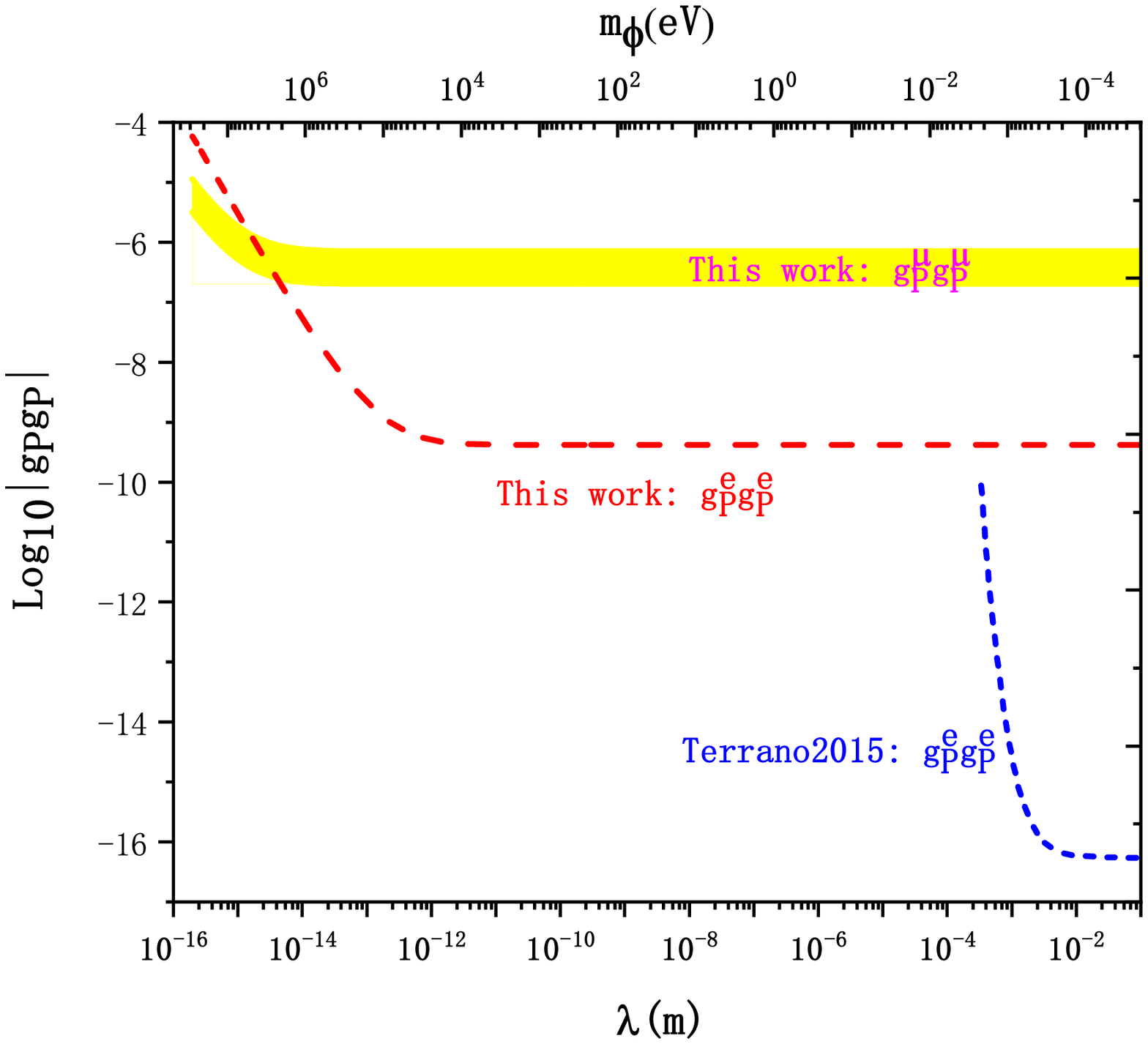}
\centerlast{\small{FIG.3 (Color online) Constraint as as a function of the interaction range $\lambda$ (ALP mass). The (yellow) shaded region is the 2-$\sigma$ band for $|g_P^{\mu}g_P^{\mu}|$ and the (red) long-dashed line is the
constraint to $|g^e_{S}g^e_{P}|$.  The (blue) short-dashed line is the result of Ref.\cite{TER15}. }}
\end{figure}
\begin{figure}[htbp]\label{Fig.gSgP}
\centering
\includegraphics[scale=0.4, angle=0]{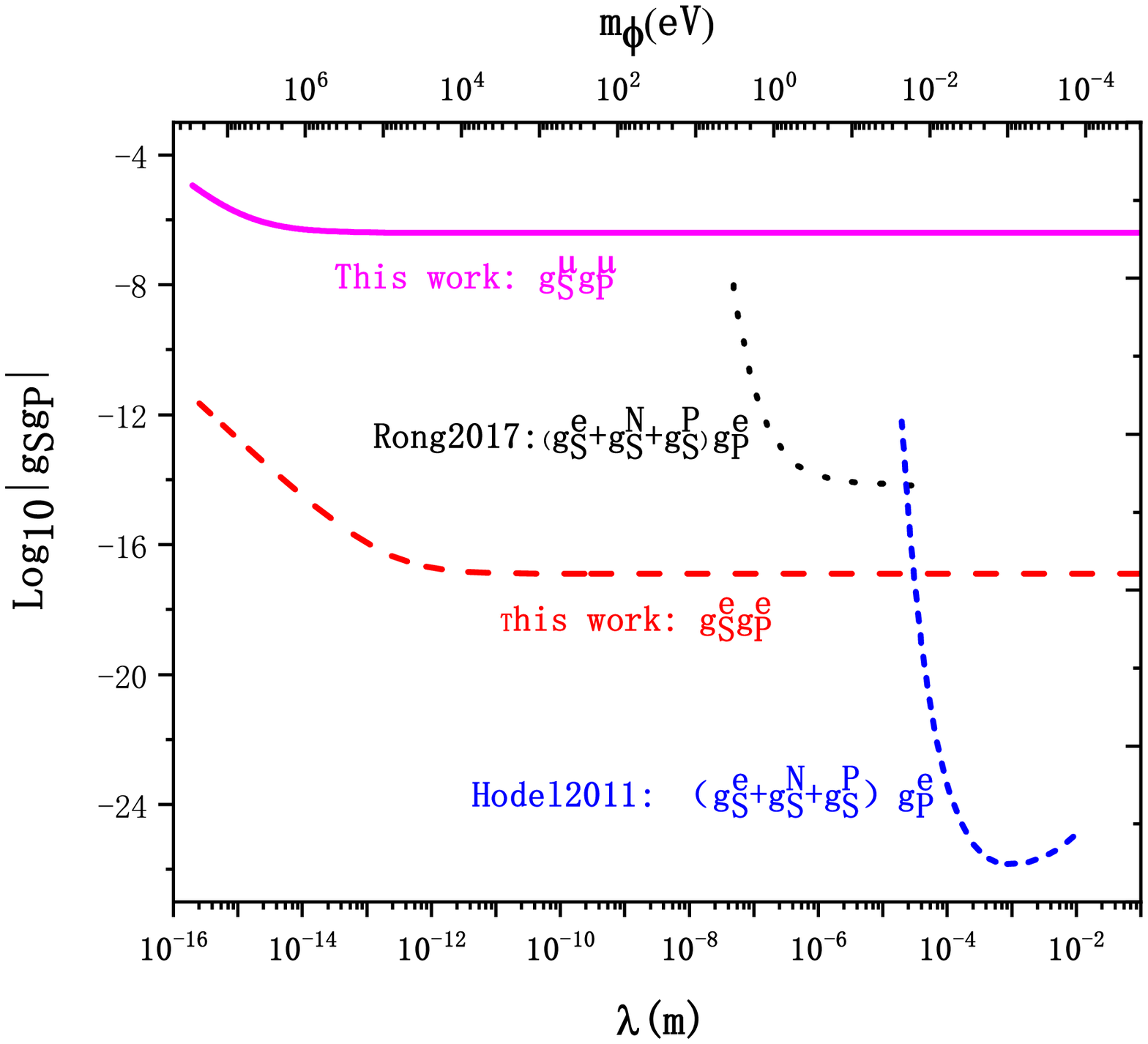}
\centerlast{\small{FIG.4 (Color online) Constraint as as a function of the interaction range $\lambda$ (ALP mass). The (magenta) solid line is the constraint for $|g_S^{\mu}g_P^{\mu}|$ . The long dashed line(red) is for $|g_S^eg_P^e|$. The (blue) dashed line is the result of Ref.\cite{HOE11}, and the (black) short-dashed line is the result from Ref.\cite{RON17}. }}
\end{figure}
We also apply this method to the electron for comparisons. Theoretically, the anomalous magnetic moment and EDM of the electron can be predicted by the Standard Model \cite{AOY17,AOY15,AOY12}. Combined with the best known experimental measurement given in Ref.\cite{ACM18}, the possible contributions due to new physics beyond the standard model are found to be:
\begin{equation}
|a_e^{exp}-a_e^{SM}|<2.66\times10^{-12}, 95\% C.L.
\end{equation}
\begin{eqnarray}
|d_e|<1.2\times10^{-29}e.cm, 95\% C.L.
\end{eqnarray}
 Subsequently, corresponding constraints for electrons can be established, shown by the (red) dashed lines in Figs.2, 3 and 4. One can see that this method could work for the ranges from $\sim$nm to $\mu$m. It gives a constraint $\sim$3 orders more stringent than a recent experimental work \cite{RON17} at $\mu$m ranges.  At long interaction ranges($m_\phi <\sim$keV), using data from atomic and molecular EDM experiments\cite{DZU18}, Ref.\cite{STA18} gives a constraint on $g_S^eg_P^e$ $\sim$3 times more stringent than the result presented Fig.4. At short ranges($m_\phi >\sim$MeV), the loop induced EDM dominates thus the method presented in this work might work better\cite{STA18T}. We emphasize here that we applied the method to electron mainly for comparison purpose, and muons are the main target for this work.
 Since the mass-sources used experimentally usually contain protons, neutrons and electrons, it is not easy to obtain the pure constraints for interactions only between electrons. Many present experiments actually gave constraints for $(g_S^e+g_S^P+g_S^N)g_P^e$. By using the anomalous magnetic moment and EDM, we can indeed obtain constraints for interactions only between fundamental fermions, as shown in Figs.2,3 and 4.
\section{conclusion and discussion}
The 3.7$\sigma$ difference between the theoretical prediction and the experimental measurement of the anomalous magnetic moment of muons, can be explained by the long-range new interactions mediated by ALPs. The long-range muonic new interaction could be either Monopole-Monopole (originated from the SS coupling) or Dipole-Dipole type (originated from the PP coupling), as shown in Fig.2 and 3 respectively. We also constrain the Monopole-Dipole type interaction (originated from the SP) for muons, as shown in Fig.4.

Previously, many experiments and studies have been performed for searching ALP-mediated new interactions associated with electrons, protons and neutrons only. Although short-range muonic forces have been proposed to solve the proton radius puzzle, studies considering the long-range muonic interactions mediated by ALPs have not been conducted yet, according to our best knowledge. The charge radius puzzles of the muonic hydrogen and deuteron, and  the anomalous magnetic moment of muons suggest that there might be new physics or new interactions related to muons.

It would be fascinating to include muons into the searching business for the long-range new interactions mediated by ALPs. Low energy muon beams with 100$\%$ polarization are frequently used in condensed matter physics \cite{AND05} and fundamental physics \cite{BAL16}. If a nonmagnetic mass-source can be put in the region close to muon beams, using the scheme as suggested in Ref.\cite{YAN14}, constraints on $(g_S^e+g_S^P+g_S^N)g_P^\mu$ at long distance can be obtained by measuring changes of the muon polarization. Furthermore, if polarized electron spin-density sources, as the $\mu$-metal shielded SmCo$_5$ \cite{TER15,JI18}, are used in experiments, then constraints on $g_P^eg_P^\mu$  can be established.  It is not hard to imagine that the muonic new interactions mediated by light vector particles can also be searched for using experimental schemes as in Ref.\cite{YAN14}.

Using the same method, we can also give constraints of $g_S^eg_S^e$, $g_P^eg_P^e$ and $g_S^eg_P^e$ for electrons. This method works at small distances from $\sim$nm to $\sim\mu$m. Moreover, it can give pure constraints only between the electrons, while many other methods 
cannot easily isolate the contributions from other fermions like protons or neutrons. Our results for the electron are consistent with zero. This fact indicates the muon might be a more interesting target for these new interactions.

\begin{acknowledgements}
 We acknowledge support from the National Natural Science Foundation of China, under grant 91636103, 11675152,11875238.  This work was also supported by National Key Program for Research and Development(Grant 2016YFA0401500 /1/2/3/4). We thank Dr. Queiroz and Dr. Stadnik for providing us useful references. We thank Dr. Stadnik for helpful discussions.
 \end{acknowledgements}


\begin{thebibliography}{10}
\expandafter\ifx\csname natexlab\endcsname\relax\def\natexlab#1{#1}\fi
\expandafter\ifx\csname bibnamefont\endcsname\relax
  \def\bibnamefont#1{#1}\fi
\expandafter\ifx\csname bibfnamefont\endcsname\relax
  \def\bibfnamefont#1{#1}\fi
\expandafter\ifx\csname citenamefont\endcsname\relax
  \def\citenamefont#1{#1}\fi
\expandafter\ifx\csname url\endcsname\relax
  \def\url#1{\texttt{#1}}\fi
\expandafter\ifx\csname urlprefix\endcsname\relax\def\urlprefix{URL }\fi
\providecommand{\bibinfo}[2]{#2}
\providecommand{\eprint}[2][]{\url{#2}}
\bibitem{PDG16} C.P. Patrignani {\it et al.} (Particle Data Group),{ Chinese. Phys. C},{\bf 40},{100001},(2016).
\bibitem{FU17} C.Fu {\it et al.}, Phys. Rev. Lett. {\bf 119}, 181806(2017).
\bibitem{KIM10} J.E. Kim, and G. Carosi, Rev. Mod. Phys. {\bf 82}, 557(2010).
\bibitem{PEC77}R.D. Peccei and H.R. Quinn, Phys. Rev. Lett. {\bf 38}, 1440(1977).
\bibitem{MOO84} J.E. Moody and F. Wilczek, Phys. Rev. D {\bf 30}, 130(1984).
\bibitem{PEC06}R.D. Peccei,arXiv:hep-ph/0607268,(2006).
\bibitem{MAN14}S. Mantry, M. Pitschmann, and M.J. Ramsey-Musolf, Phys. Rev. D {\bf 90}, 054016 (2014) .
\bibitem{VAS09} G. Vasilakis, J.M. Brown, T.W. Kornack, W. Ketter, M.V. Romalis, Phys. Rev. Lett. {\bf 104}, 261801(2009).
\bibitem{HOE11}S.A.Hoedl, F.Fleischer, E.G.Adelberger, B.R.Heckel, Phys. Rev. Lett. {\bf 106}, 041801(2011). 
\bibitem{RAF12}G.Raffelt Phys. Rev. D {\bf 86}, 015001 (2012) . 
\bibitem{BUL13} M. Bulatowicz {\it et al.},  Phys. Rev. Lett. {\bf 111}, 102001(2013).
\bibitem{LED13} M. P. Ledbetter, M.V. Romalis, and D. F. J. Kimball Phys. Rev. Lett. {\bf 110}, 040402(2013).
\bibitem{TUL13} K. Tullney {\it et al.}, Phys. Rev. Lett. {\bf 111}, 100801(2013).
\bibitem{LES14} T. M. Leslie, E.Weisman, R. Khatiwada, and J. C. Long, Phys. Rev. D {\bf 89}, 114022 (2014)
\bibitem{YAN14}H. Yan,  {\it et al.}, Eur. Phys. J. C  {\bf74} 3088(2014).
\bibitem{CHU15}P.-H. Chu E.Weisman, C.-Y. Liu and J.C.Long, Phys. Rev. D {\bf 91}, 102006 (2015) .
\bibitem{KOT15}S.Kotler, R.Ozeri, D.F.J. Kimball,  Phys. Rev. Lett. {\bf 115}, 081801(2015). 
\bibitem{TER15} W. A. Terrano, E. G. Adelberger, J. G. Lee, and B. R. Heckel, Phys. Rev. Lett. {\bf 115}, 201801(2015).
\bibitem{CHU16}P.-H. Chu Y.J.Kim, C and I.Savukov, Phys. Rev. D {\bf 94}, 036002 (2016) .
\bibitem{FIC17}F.Ficek, D.F.J. Kimball, M.G. Kozlov, N.Leefer, S.Pustelny,and D.Budker, Phys. Rev. A {\bf 95}, 032505(2017) .
\bibitem{DOB06} B. Dobrescu and I. Mocioiu, J. High Energy Phys. {\bf 11}, 005 (2006).
\bibitem{ANT13}A. Antognini,  {\it et al.}, SCIENCE, {\bf 339} 417(2013)
\bibitem{POH16}R. Pohl,  {\it et al.}, SCIENCE, {\bf 353} 669(2016)
\bibitem{BAT11}B. Batell,  D. McKeen, and M. Pospelov,  Phys. Rev. Lett. {\bf 107}, 011803(2011).
\bibitem{BAR12}V. Barger, C.W. Chiang, W.Y. Keung, and D. Marfatia, Phys. Rev. Lett. {\bf 108}, 081802(2012).
\bibitem{BER91}W.Bernreuther, M.Suzuki, Rev. Mod. Phys. {\bf 63}, 313, (1991). 
\bibitem{ITZ80}C. Itzykson and J.-B. Zuber, Quantum Field Theory, McGraw-Hill, Inc. (1980)
\bibitem{ZEE10}A. Zee, Quantum Field Theory in a Nutshell, Princeton University Press $\cdot$ Princeton and Oxford (2010)
\bibitem{SUPL}Supplementary Material for "Constraining New Muonic  Interactions Meditated by Axion-Like-Particles",which includes Refs. \cite{BER91,ITZ80,ZEE10}.
\bibitem{LIN16}M. Lindner, M. Platscher and F.S. Queiroz, arXiv:1610.06587,(2016).
\bibitem{BLU18}T.Blum et al. (RBC and UKQCD Bollaborations),  Phys. Rev. Lett. {\bf 121}, 022003 (2018)
\bibitem{BEN09} G. W. Bennett et al. (Muon (g-2) Collaboration), Phys. Rev. D {\bf 80}, 052008 (2009) 
\bibitem{AOY17} T. Aoyama, M. Hayakawa, T. Kinoshita, and M. Nio, Phys. Rev. D {\bf 96}, 011901(E) (2017) 
\bibitem{AOY15} T. Aoyama, M. Hayakawa, T. Kinoshita, and M. Nio, Phys. Rev. D {\bf 91}, 033006 (2015) 
\bibitem{AOY12} T. Aoyama, M. Hayakawa, T. Kinoshita, and M. Nio, Phys. Rev. Lett. {\bf 109}, 111807 (2012)
\bibitem{ACM18}The ACME collaboration, NATURE {\bf 562}, 355-360(2018)
\bibitem{BRO14}K.R.Brown, SCIENCE {\bf 343}, 255-256(2014)
\bibitem{RON17}X.Rong  {\it et al.}, Nature Communications, {\bf 9},739,(2018).
\bibitem{STA18}Y.V. Stadnik, V.A. Dzuba, and V.V. Flambaum, Phys. Rev. Lett. {\bf 120}, 013202(2018).
\bibitem{DZU18}V.A. Dzuba, V.V. Flambaum, I.B. Samsonov, and Y.V. Stadnik,   Phys. Rev. D {\bf 98}, 035048(2018).
\bibitem{STA18T}Y.V. Stadnik, private communication.
\bibitem{ADE09} E. G. Adelberger, J.H. Gundlach, B.R. Heckel, S. Hoedl, S. Schlamminger,  Prog. Part. Nucl. Phys. {\bf 62}, 102 (2009).
\bibitem{HUD11}J.J.Hudson, D.M.Kara, I.J.Smallman, B.E.Sauer, M.R.Tarbutt \& E.A.Hinds, NATURE, {\bf 473}, 493(2011).

\bibitem{AND05}D. Andreica, Encyclopedia of Condensed Matter Physics, (2005)
\bibitem{BAL16}A.M. Baldini et al (The MEG Collaboration), Eur. Phys. J. C  , {\bf 76},223 (2016)
\bibitem{JI18}W.Ji  {\it et al.}, arXiv:1803.02813


\end{thebibliography}
\end{document}